\documentclass[
 reprint,
superscriptaddress,
  aps,
superscriptaddress,
 amsmath,amssymb,
 longbibliography
]{revtex4-1}

\usepackage{graphicx}
\usepackage{dcolumn}
\usepackage{bm}
\usepackage{amsmath,amssymb,amstext,amsthm}
\usepackage{braket}
\usepackage{bbold}
\usepackage{bbm}
\usepackage{xcolor}
\usepackage{ gensymb }
\usepackage{soul}
\usepackage{subcaption}
\usepackage{mathrsfs}
\usepackage{svg}

\newcommand{\pendo}{pendell\"osung }

\begin{document}

\setlength{\abovedisplayskip}{1pt}

\title{Quantum Information Approach to the Implementation of a Neutron Cavity}

\author{O. Nahman-L\'{e}vesque} 
\affiliation{Institute for Quantum Computing, University of Waterloo,  Waterloo, ON, Canada, N2L3G1}
\affiliation{Department of Physics, University of Waterloo, Waterloo, ON, Canada, N2L3G1}

\author{D. Sarenac}
\affiliation{Institute for Quantum Computing, University of Waterloo,  Waterloo, ON, Canada, N2L3G1}

\author{D. G. Cory}
\affiliation{Institute for Quantum Computing, University of Waterloo,  Waterloo, ON, Canada, N2L3G1}
\affiliation{Department of Chemistry, University of Waterloo, Waterloo, ON, Canada, N2L3G1}

\author{M. G. Huber}
\affiliation{National Institute of Standards and Technology, Gaithersburg, Maryland 20899, USA}

\author{D. A. Pushin}
\email{dmitry.pushin@uwaterloo.ca}
\affiliation{Institute for Quantum Computing, University of Waterloo,  Waterloo, ON, Canada, N2L3G1}
\affiliation{Department of Physics, University of Waterloo, Waterloo, ON, Canada, N2L3G1}

\date{\today}

\begin{abstract}

Using the quantum information model of dynamical diffraction we consider a neutron cavity composed of two perfect crystal silicon blades capable of containing the neutron wavefunction. We show that the internal confinement of the neutrons through Bragg diffraction can be modelled by a quantum random walk. Good agreement is found between the simulation and the experimental implementation. Analysis of the standing neutron waves is presented in regards to the crystal geometry and parameters; and the conditions required for well-defined bounces are derived. The presented results enable new approaches to studying the setups utilizing neutron confinement, such as the experiments to measure neutron magnetic and electric dipole moments. 

\end{abstract}

\pacs{Valid PACS appear here}


\maketitle

\section{Introduction}

The advancement of silicon fabrication techniques in the 20\textsuperscript{th} century has enabled the use of perfect crystals in neutron science to resolve dynamical diffraction effects. The most successful application was perfect-crystal neutron interferometry where three well-polished blades from a single ingot of silicon achieve centimeter-scale path separation and coherent superposition~\cite{rauch1974test, Klepp_2014, rauch2015neutron, pushin2015neutron, huber2019overview}. Hallmark experiments with a perfect-crystal neutron interferometer (NI) include the first demonstration of gravity on a quantum particle~\cite{COW1975}, probing of dark energy/fifth forces~\cite{lemmel2015neutron,li2016neutron}, observing the 4$\pi$ symmetry of spinor rotation~\cite{rauch1975verification}, and observing neutron orbital angular momentum~\cite{clark2015controlling,sarenac2016holography}.

The theory of dynamical diffraction (DD) models the neutron propagation through perfect crystals. Albeit powerful, it becomes impractical when considering complicated geometries. For example, neutron cavities that employ DD have been experimentally investigated for use as a neutron storage device~\cite{jericha1996performance,jericha1997cold,jericha2000neutron,jaekel2005new,jaekel2005new,rauch2001quantum,facchi2003optimization}. 
However, in-depth modelling of the neutron propagation through the cavity and parameter analysis is lacking.

The recently introduced quantum information (QI) model of DD attempts to simplify the mathematics by treating the neutron propagation as a quantum random walk~\cite{Nsofini_2016,nsofini2019coherence,nsofini2017noise,nahman2022generalizing}. Good agreement has been found between the model and the standard DD effects such as the Borrmann triangle, \pendo oscillations, and the spatial intensity profiles of the NI paths.

Here we apply the QI model of DD to the case of a neutron cavity built from two perfect crystal silicon blades. We show that such an arrangement can contain the neutron wavefunction. The behaviour of the neutron modes inside the cavity is characterized with respect to the cavity geometry, and good agreement is found between an initial experimental implementation and the model. In addition to providing new insights into neutron storage devices, the presented work lays the foundation for simulating the proposed measurements of the neutron magnetic and electric dipole moment~\cite{dombeck2001measurement,gentile2019direct}.

\section{QI model for dynamical diffraction}
\label{QImodel}

When considering neutron diffraction through crystals, DD effects must be considered. However, the standard theory of DD is limited to idealized perfect crystals (no impurities or defects)  and practically limited to simple geometries. In \cite{Nsofini_2016}, it was demonstrated that there exists a simpler, more computationally accessible QI based model in terms of a quantum random walk. In the QI model, the crystal is represented as a lattice of nodes which act as a unitary operator on the neutron state by either transmitting or reflecting it to its nearest neighbours. The neutron input state to every node is given by a two-state vector:

\begin{equation}
    \alpha \ket{a}+ \beta \ket{b}
    \quad\mathrm{or}\quad 
    \begin{pmatrix}
           \alpha \\
           \beta \\
    \end{pmatrix}
\end{equation}

where $\ket{a}$ and $\ket{b}$ are the states propagating onto the node in the up and down directions, respectively. The action of one node is given by:

\begin{equation}
    U_i = \ket{a_{i+1}}(t_a \bra{a_i} + r_b \bra{b_i})+\ket{b_{i-1}}(r_a \bra{a_i}+t_b\bra{b_i})
    \label{crystalUnitary}
\end{equation}

\noindent
with coefficients
\begin{equation}
\begin{split}
    t_a = e^{i\xi}\cos{\gamma},\;\; r_b = e^{i\zeta}\sin{\gamma}\\
    r_a = -e^{-i\zeta}\sin{\gamma},\;\; t_b = e^{-i\xi}\cos{\gamma}
    \label{eq:coefficients}
\end{split}
\end{equation}

\noindent The parameter $\gamma$ controls the amplitude of reflection/transmission at each node, while $\xi$ and $\zeta$ are phase parameters.

\begin{figure}
    \centering
    \includegraphics[width=\linewidth]{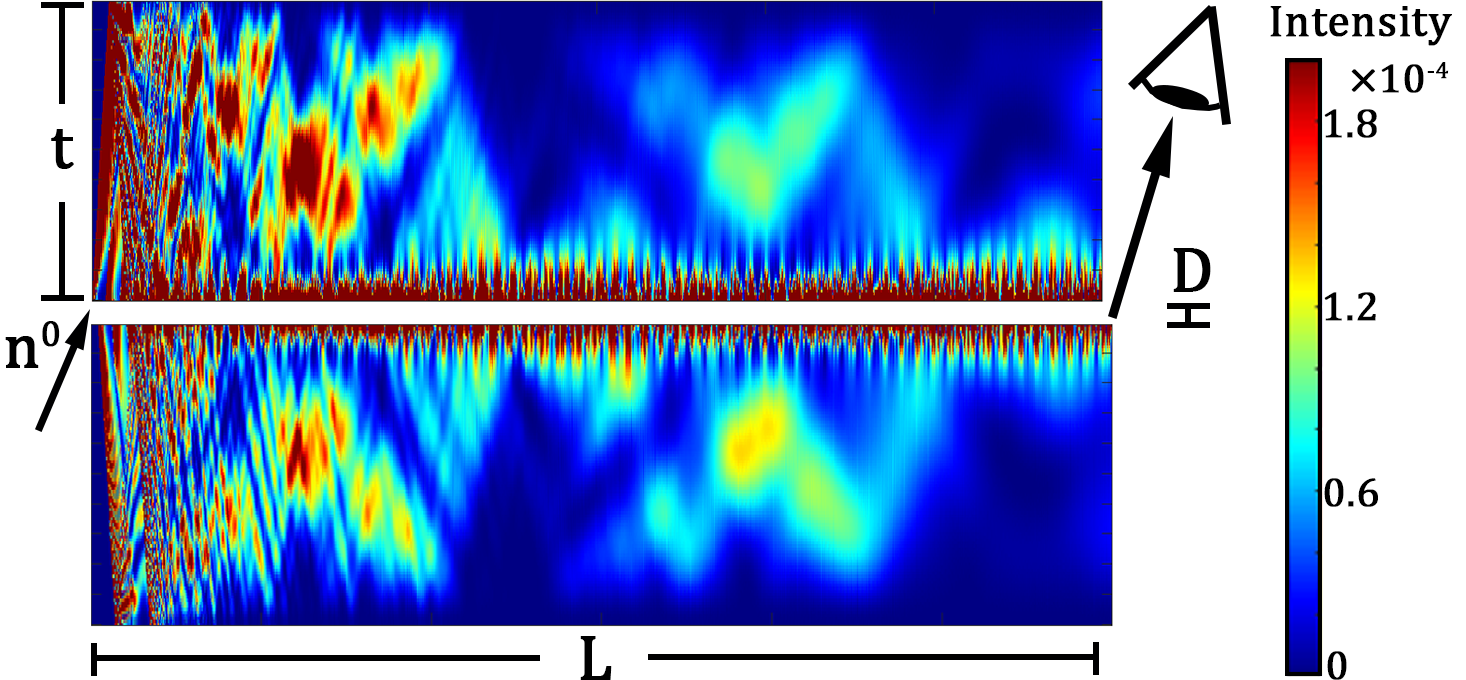}
    \caption{The simulated neutron intensity inside two perfect-crystal silicon blades acting as a neutron cavity. The neutron is initialized as an incident spherical wave to the corner of the top crystal. Given that most of the beam gets initially transmitted, the range of the color bar is limited to emphasize the confined intensity. The crystals are placed a distance $D$ apart, have thickness $t$ and length $L$. After the first few bounces, most of the leftover neutron intensity is contained within a small band on the inside edges of the crystals. This intensity is preserved with further propagation. A detector can be placed at the exit of the crystals to capture the neutrons exiting the cavity. This simulation ($\Delta_H \approx 50 \mu$m, $t \approx 0.5$mm, $D \approx 0.2$mm, $L \approx 3$cm) considers a very thin crystal slice to better illustrate the behavior near the gap.}
    \label{fig:bragg_loop}
\end{figure}

The full state vector $\psi$ at every layer of a simulation of height $h$ is represented by a size $2h$ column vector, where the entries $m,m+1$ are the up and down inputs to node number $\frac{m+1}{2}$. Every lattice column $c_i$ has a corresponding unitary operator $C_i$, and the state vector after $n$ layers is given by
\begin{equation}
    \psi_n = \prod_{i=0}^{n-1} C_{n-i} \psi_0
\end{equation}
In \cite{nahman2022generalizing}, it was shown that there is an equivalence between the simulation parameters and a given experimental configuration in both the Laue and Bragg geometry, given by
\begin{equation}
    n \cdot \gamma = \frac{\pi d}{2\Delta_H}
    \label{mainrelation}
\end{equation}
where $n$ is the number of simulation layers needed to simulate a real-space distance $d$ with parameter $\gamma$. $\Delta_H$ is the \pendo period inside the crystal, and is given by

\begin{equation}
    \Delta_H = \frac{\pi V_{cell}\cos\theta_B}{\lambda |F_H|}
    \label{eq:Pend}
\end{equation}
where $V_{cell}$ is the volume of a crystal unit cell, $\theta_B$ is the Bragg angle, $\lambda$ is the neutron wavelength and $F_H$ is the crystal structure factor. For a material such as silicon, with common wavelengths used in experiments, $\Delta_H$ is on the order of $50~\mu$m.

\section{Modelling of a Neutron Cavity}

\begin{figure}
    \centering
    \includegraphics[width = \linewidth]{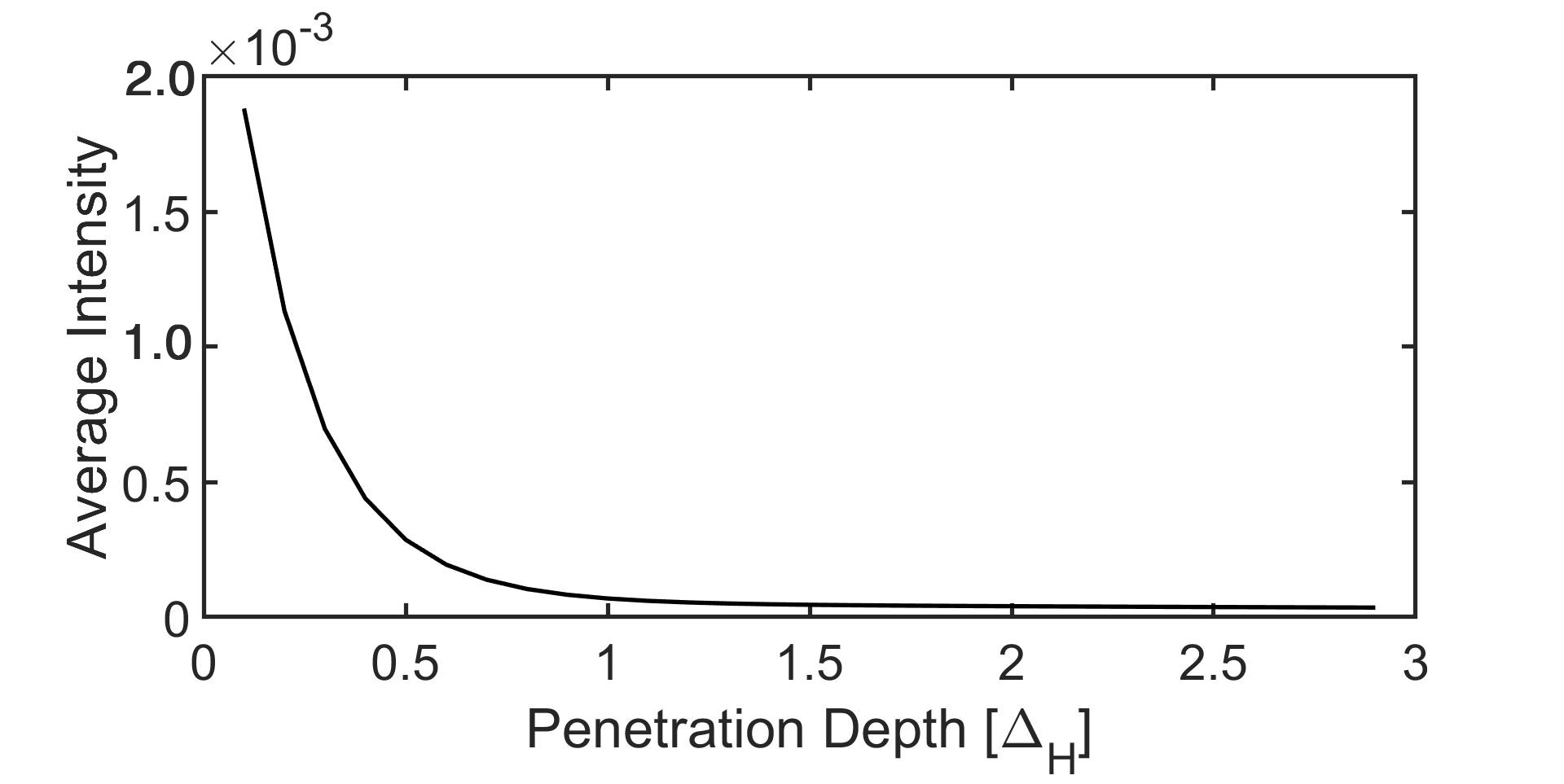}
    \caption{The average neutron intensity across a horizontal slice of the top crystal shown in Fig.~\ref{fig:bragg_loop} is plotted as a function of the penetration depth inside the top crystal. Note that the starting point was at bounce 110 in order to avoid the initial high intensity region. The intensity drops sharply within the first \pendo period, indicating that the neutron is confined within a band of width $\approx\Delta_H$ inside the cavity.}
    \label{fig:intvsdepth}
\end{figure}

Dynamic diffraction theory predicts that in the Bragg geometry, neutrons falling within a narrow range of momentum centered around the Bragg condition (the Darwin width) are reflected with close to 100\% probability. In a neutron cavity composed of two perfect crystals, neutrons outside the Darwin width will escape through the crystals within the first few bounces. Conversely, neutrons inside this width are effectively confined, allowing for a great number of bounces. Interestingly, the QI model predicts that this result can be explained purely as a consequence of a quantum random walk: the total neutron wavefunction is expressed as a sum over all the possible paths through the Bragg lattice. The difference in phase accumulated by the different paths results in constructive interference only for the paths which stay confined in a small region close to the gap. 

To model such a geometry, we use a single column matrix where the nodes outside (inside) the crystals are free space (Bragg diffracting) operators. The operator for the free space nodes is given by the matrix

\begin{equation}
    U_{\textrm{free}} = \ket{a_{i+1}}\bra{a_i}+\ket{b_{i-1}}\bra{b_i}
\end{equation}

while the operator for the Bragg diffracting nodes are given by Eqn. \ref{crystalUnitary}.
As shown in Fig.~\ref{fig:bragg_loop}, without any information about transverse momentum, boundary conditions, or incident angle, the model predicts that only after a few bounces, the intensity is almost entirely localized in a narrow band around the inter-crystal gap. Most neutrons that do not settle within this band are either transmitted straight through the top of the crystal interferometer, or bounce at most once and transmit through the bottom. 


\begin{figure*}
    \centering
    \includegraphics[width = 1\textwidth]{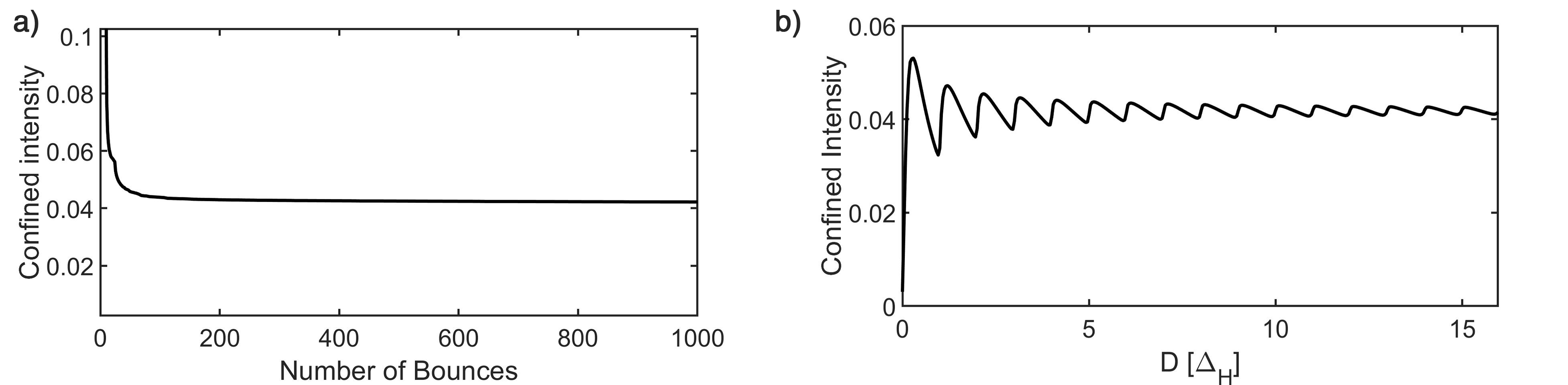}
    \caption{a) The neutron intensity remaining inside the cavity as a function of the number of bounces, for a cavity with $D = 12\Delta_H$ and $t = 87.5 \Delta_H$. The intensity drops very sharply during the first few bounces then stabilizes as the direct beam leaves the top crystal. A value for the reflectivity of the crystal of $\approx 1-1.6\times 10^{-5}$ was extracted by fitting to an inverse exponential between bounces 500 and 800. b) The neutron intensity remaining inside the cavity after a length of 16000 $\Delta_H$. The neutron undergoes a number of bounces proportional to $D^{-1}$. The final intensity in the cavity oscillates with period $D=\Delta_H$.}
    \label{composite}
\end{figure*}
Shown in Fig.~\ref{fig:intvsdepth} is the neutron average intensity across a horizontal slice of the top crystal as a function of penetration depth into the top crystal. The intensity drops sharply within the first $\Delta_H$, showing that the confined neutrons do not penetrate very deeply inside the crystal. 

\subsection{Confined Intensity}
With the QI model we can examine the neutrons which remain confined within the cavity. Fig.~\ref{composite} a) shows the confined intensity inside the cavity as a function of the number of bounces, for cavity parameters $D = 12\Delta_H$ and $t = 87.5 \Delta_H$. It is to be noted that as the neutron progresses through the cavity, the individual bounces become hard to resolve, and therefore here we are considering a length of crystal corresponding to a geometrical path of 1000 bounces. The intensity drops sharply as the direct beam transmits straight through the first crystal, and then drops slightly less at each subsequent reflection. Eventually, it settles to an effectively constant value of around $4 \,\%$ of the incoming intensity. An estimation for the reflectivity of the crystal as the number of bounces increases can be extracted from this curve, by fitting it to an inverse exponential. The reflectivity in the ``stable'' region between 500 and 800 bounces is found to be $(\approx 1-1.6\times10^{-5})$. 

In Fig. \ref{composite}b, the intensity remaining in the cavity after a length $L = 16000\Delta_H$ is plotted as a function of $D$. The confined intensity oscillates with period $D=\Delta_H$. The QI model indicates that to maximize the number of bounces for a given set of crystals, the spacing can be made very small without losing much intensity at the exit, but should be made no smaller than $\Delta_H/4$.

\begin{figure*}
    \centering
    \includegraphics[width = 1\textwidth]{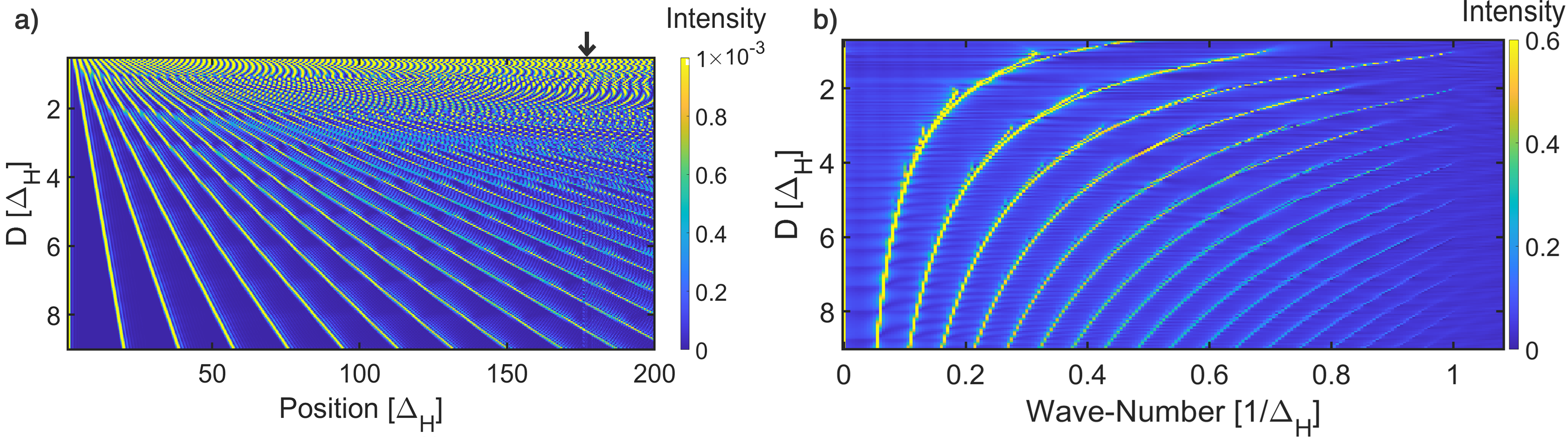}
    \caption{a) The intensity reflected from the bottom surface of the top crystal as a function of position, for different intercrystal spacing distances $D$. There are two different regimes: when the spacing is smaller than the \pendo length, the neutron state is represented by a standing wave, whose frequency depends on crystal spacing. The second regime is where the spacing is larger than the \pendo length. Here the neutron bounces are well-separated at first followed by self interference after many bounces. There is a noticeable disruption around 175 $\Delta_H$ (indicated by an downwards arrow) caused by the first reflection from the back face of the top crystal. b) The frequency spectrum of the intensity profiles shown in Fig. \ref{composite2}a. In the region of $D<\Delta_H$, there is one dominant frequency, which varies with $D$. As the spacing increases, the neutron bounces become well-separated, and new frequencies appear corresponding to the higher harmonics of the main bounce frequency.}
    \label{composite2}
\end{figure*}

\subsection{Cavity Modes}
Simulation of the neutron cavity using the QI model shows that the neutrons inside the cavity settle in one of two regimes, depending on the intercrystal distance $D$.  It should be noted that in the limit where $D \to 0$, the system reduces to a simple Laue crystal of thickness $L$, in which case the neutron will diverge to the edges of the Bormann triangle as predicted by DD theory. However, for nonzero spacing, after the first few bounces, most of the neutron current is confined within a small band of size of order $\Delta_H$ on the inside surface of the cavity, and bounces back and forth without penetrating very deeply into the crystal. Fig.~\ref{composite2}a shows the simulated reflected intensity on the surface of the top crystal throughout the cavity, for different values of $D$. For the D $<\Delta_H$ regime,
the neutron behaves like a standing wave inside the cavity, with a period dependant on $D$. For the regime with larger values of $D$ ($>2\Delta_H$), the bounces are well-separated and localized at first.  As the number of bounces increases, the neutron wave packets spread and induce interference effects. This behavior is shown in Fig.~\ref{composite2}b: for small $D$, there is one dominant frequency which changes with $D$. As $D$ increases, more pairs of harmonics are introduced.

\section{Experimental Implementation}

\begin{figure*}
    \makebox[\linewidth]{
        \includegraphics[width=\textwidth]{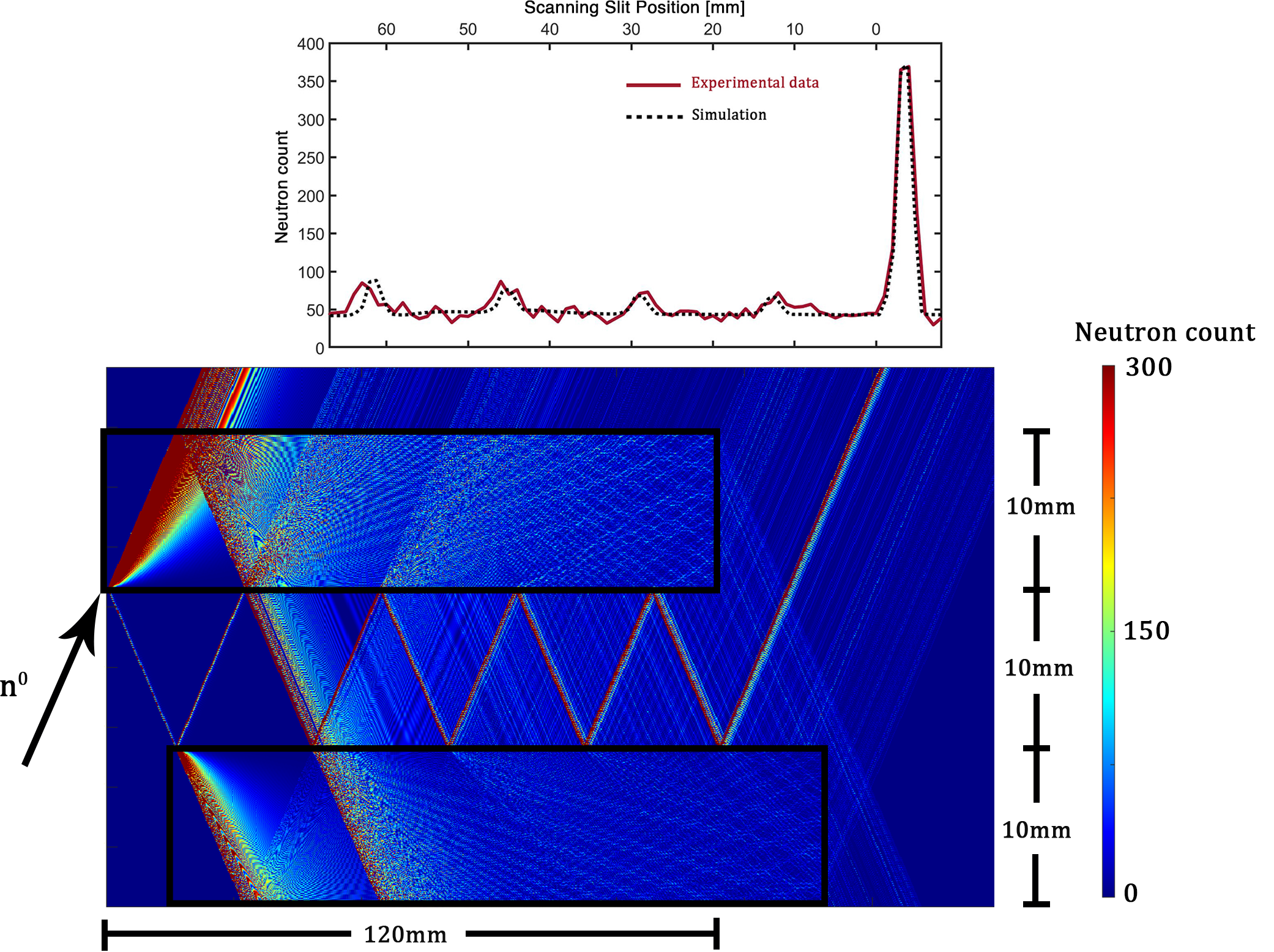}
    }
    \caption{The experimental implementation of a neutron cavity using two perfect-crystal silicon blades and a position sensitive neutron detector above the top blade. Using the described QI model of DD we can simulate the neutron propagation through the cavity. Here the nodes outside the two crystals act as identity matrices, while a quantum random walk occurs inside the two crystals. No further physics or boundary conditions are present. The geometric trajectory of the neutron as it bounces back and forth is clearly visible between the two crystals. Shown above in red is the experimental measurement corresponding to this geometry. Overlaid in the black, dashed line, is the intensity obtained via the QI model. Note that as shown in Ref.~\cite{nahman2022generalizing}, the QI model accounts for experimental parameters by convolving the intensity output with the shape of the exiting beam. Good agreement is found between the simulation and experiment.}
    \label{fig:MDMsim}
\end{figure*}

An experimental implementation of a neutron cavity was performed at the NIOF beamline at the National Institute of Standards and Technology (NIST) center for neutron research (NCNR). The setup is shown in Fig.~\ref{fig:MDMsim}.  A beam of neutrons with wavelength 0.235 nm was propagated through a 10 mm wide silicon neutron cavity composed of two 10 mm thick perfect-crystal silicon blades (220 reflection) attached to a common base. The theoretical \pendo length for this configuration is found to be $50.38$ $\mu$m. The neutrons underwent four well-defined bounces before leaving the cavity at the exit. A scanning slit and an integrating detector were placed behind the top crystal, and were used to map the spatial intensity of neutrons along the crystal cavity. 

Using the QI model we can simulate the experimental configuration. The simulation parameters were chosen according to the equivalence relation of  Eq.~\ref{mainrelation}. As shown in Ref.~\cite{nahman2022generalizing} we can account for the experimental factors such as the spread of the incident beam, beam divergence, and slit size through the analysis of a Bragg diffraction peak. Similar to the methods of Ref.~\cite{nahman2022generalizing} the intensity penetrating through the top crystal was convolved with the measured shape of the exit peak. The simulated intensity throughout the setup is shown as a false-color map, and the simulated integrated intensity at the detector is shown above the setup in a black, dashed line. The simulated intensity profile is in good agreement with the measured intensity profile. It can be observed that the majority of the contribution to the exiting beam of the cavity comes from the classical bouncing path.

\section{Conclusion and Discussion}

In conclusion, we have applied the QI model of dynamical diffraction to the case of a neutron cavity composed of two perfect crystal silicon blades. The model enables us to study the dependence of the cavity modes on the system geometry. Full analysis is provided and good agreement is found between an experimental implementation and the simulation of the particular setup.

Several exciting applications and proposed measurements rely on neutron confinement inside such cavities. Our work presents an in-depth analysis of the neutron behaviour inside the crystal blades and the crystal spacing. Therefore, we expect that the QI model will become a standard and easily-accessible tool used for experimental analysis that relies on neutron dynamical diffraction.


\section{Acknowledgements}

This work was supported by the Canadian Excellence Research Chairs (CERC) program, the Natural Sciences and Engineering Research Council of Canada (NSERC) Discovery program, the Collaborative Research and Training Experience (CREATE) program, the Canada  First  Research  Excellence  Fund  (CFREF), and the US Department of Energy, Office of Nuclear Physics, under Interagency Agreement 89243019SSC000025. The authors would like to thank Ansel Himmer and Tim Olsen for helping collect the experimental data.

\bibliography{refs}

\end{document}